%% file: main.tex
\documentclass[preprint,12pt]{elsarticle}

\usepackage{amssymb}
\usepackage{makecell}

\journal{Future Generation Computer Systems}

\begin{document}

\begin{frontmatter}



\title{An Edge-based Architecture to Support the Execution of Ambience Intelligence Tasks using the IoP Paradigm}


\author[a]{Khaled Alanezi}
\author[b]{Shivakant Mishra}

\address[a]{Department of Computing, College of Basic Education, PAAET, Kuwait}
\address[b]{Computer Science Department, University of Colorado, Boulder, USA}

\input{abstract}




\end{frontmatter}


\input{introduction}

\input{related}

\input{design}

\input{protocol}

\input{implementation}

\input{conclusion}





\bibliographystyle{elsarticle-num}
\bibliography{references}

\end{document}

%% file: abstract.tex
\begin{abstract}

In an IoP environment, edge computing has been proposed to address the problems of resource limitations of edge devices such as smartphones as well as the high-latency, user privacy exposure and network bottleneck that the cloud computing platform solutions incur.
This paper presents a context management framework comprised of sensors, mobile devices such as smartphones and an edge server to enable high performance, context-aware computing at the edge. 
Key features of this architecture include energy-efficient discovery of available sensors
and edge services for the client, an automated mechanism for task planning and execution 
on the edge server, and a dynamic environment where new sensors and services may be added to the framework.
A prototype of this architecture has been implemented, and an experimental evaluation using
two computer vision tasks as example services is presented. Performance measurement shows that
the execution of the example tasks performs quite well and the proposed framework is well suited for an edge-computing environment. 
\end{abstract}

%% file: introduction.tex
\section{Introduction}
\label{sec:introduction}
Bringing context-aware computing to mobile smartphones promises to unlock a range of interesting applications from smart personal digital assistants to accurate health monitoring and contextualized ads. However, mobile technology is still far from fully achieving this vision as many challenges to provide accurate and efficient context-aware computing on mobile devices remain unsolved. First, the data collected by mobile devices often suffer from noise that leads to huge degradation of the classifier accuracy. Take for example a mobile device with a camera that is not fully facing the object to perform object recognition or a microphone that is somewhat far from an important audio context (e.g. laughter \cite{hagerer2017did} or a cough \cite{larson2011accurate}). Also, mobile devices can suffer from disadvantageous positions which can also lead to inaccurate classification such as a mobile device buried inside a pocket or a purse. Second, variations in context \cite{li2016deepcham} of when and where the sample for classification is collected makes it difficult to use generic pre-trained classifiers in mobile environments. For example, a classifier trained with images taken in bright environment would perform poorly in a dark environment. Likewise, an audio classifier that is trained with high-volume audio clips would perform poorly when classifying low-volume audio clips. The third challenge when it comes to adapting context-aware computing comes from the limited energy of mobile devices. Despite continuous efforts in improving battery technology by the manufacturers of mobile devices, these devices will remain limited in their energy (and computing) capability when compared to their tethered counterparts. Consequently, it is impractical to solely depend on the mobile device to perform these tasks as this will drain the device's battery thereby negatively impacting user experience. 
To address the aforementioned challenges, researchers have looked at utilizing the abundance of resources in the cloud to train deeper (and hence more accurate) classifiers and perform the classification required for ambience intelligence tasks. More recently, researchers proposed the concept of pushing the capabilities of the cloud to the edge of the network \cite{shi2016edge}, namely edge computing, to address the problems of high-latency, user privacy exposure and network bottleneck that the cloud computing paradigm suffers from. 

Inline with the edge computing paradigm, in this work, we propose that the availability of an edge server provides a unique and an unprecedented opportunity to bring context-aware mobile computing to fruition by addressing the above stated challenges. The edge server is a server node that is installed at the edge of the network with dedicated resources to perform data processing and computation offloading for ambience intelligence tasks needed to achieve context-awareness. 
We propose to utilize the edge server as a trusted and smart coordinator between mobile devices. A mechanism is established to allow mobile devices to act on behalf of their users by registering their capabilities and negotiating the execution of ambience intelligence tasks with the edge server. Consequently, the edge server utilizes this broad information to devise execution plans to serve these tasks ensuring best-effort in terms of accuracy and energy efficiency. The design choice of allowing mobile devices to act as proxies for their owners in the environment follows the Internet of People (IoP) principle \cite{conti2018internet, miranda2015internet} seen as a better model for serving the ever expanding network edge.

In particular, in this paper we propose a framework for context-aware computing for an
edge environment for enabling context-aware IoP applications. This framework allows 
mobile clients to discover the sensors and the services provided at an edge environment
in an energy efficient manner.
In addition, it allows clients to contribute sensor data as well as new service tasks to 
be executed at the edge server. The edge server is a central coordinator in this framework
responsible for keeping an updated repository of available sensors and services, advertising 
these available sensors/services so that mobile clients can discover them and make use 
of them, and efficiently planning for service computation on client's behalf. The
framework automates the entire process of edge server maintaining an updated repository of available sensors/services, sensor/service advertisement and discovery,
execution planning and interactions with the mobile clients.
To demonstrate the efficacy of the proposed framework, we have implemented a
prototype of the proposed framework using BLE as the communication medium between clients and the edge server, WiFi as the communication medium between IoT sensors and the edge server, and object recognition and face recognition services as example ambience intelligence tasks.
Performance measurements from this prototype show that the proposed framework is well 
suited for an edge computing environment to support context-aware IoP applications.

%% file: related.tex
\section{Related Work}

This research sets at the intersection of three recent research thrusts under the umbrella of ubiquitous computing. Hence, we divide this literature review section into three main parts to place our work in proper perspective.
\subsection{The Internet of People}
The IoP paradigm \cite{conti2018internet, miranda2015internet} is an extension on top of the current Internet and Internet of Things architectures that advocate a user-centric approach for building and organizing networks on the edge. In essence, using this approach, user devices will move from being mere consumers of services to participating in self-organizing communities that act on behalf of the forming users to achieve a form of collective intelligence \cite{lagerspetz2018pervasive}. This in turn is envisioned to achieve substantial benefits from accurate sensing to faster execution times and preserving user privacy. The architecture we propose in this work builds on the same concept where mobile devices can act collectively on behalf of their users in joining coordination networks. However, a distinguishing factor for our work is the employment of an edge server to play central role in coordinating  devices. In addition to its powerful computing capabilities and physical proximity to IoT nodes, the edge server has a birds-eye-view of the IoT environment that we utilize to devise smarter collaboration plans. 

Another architecture to serve the IoP paradigm is built using cloud computing and microservices architecture \cite{macias2019microservice} to aid in the development of IoT and people applications. Our proposed solution is similarly inspired by the IoP paradigm but we employ edge computing rather than the cloud while having different objectives (i.e. accuracy and energy efficiency) in mind.

\subsection{Opportunistic Computing}
The concept of opportunistic computing \cite{conti2010opportunities} proposes that mobile devices with physical proximity can work towards, and share the burden of common tasks. A key underlying assumption here is that collaborating nodes will be interested in achieving a common goal. CoMon \cite{lee2012comon} presented a solution based on opportunistic computing with the goal of allowing nearby mobile devices to take turns in monitoring a context of shared interest. Participating devices would save energy by splitting the burden among them. We share the same goal (i.e. context monitoring) with CoMon but depend in our architecture on the edge server as the coordinator for arranging between mobile devices. Microcast \cite{keller2012microcast} also utilizes collaborations between smartphones to split up video streaming task assuming that co-located users are watching the same content. Panorama \cite{alanezi2015panorama} is another system that is based on collaborative computing. However, in addition to nearby mobile devices, Panorama considers available edge and cloud resources to further minimize the task allocated to battery powered devices. 

\subsection{Edge Computing, Edge Intelligence and Computer Vision}
Edge computing proposes to push the computing capabilities of the cloud to the edge of the network to serve delay sensitive tasks. Our architecture is inspired by this approach, but we take one step further by performing coordination through the edge server. We start by describing works that take similar approach as ours in utilizing an edge server for coordination. Privacy mediators \cite{davies2016privacy} proposed to use edge servers for coordination role. However, the primary goal is to mediate between the privacy policy of the IoT owner and the privacy preferences of mobile users in the environment. Also, DeepCham \cite{li2016deepcham} is a solution that is mediated using an edge server. The goal of DeepCham is to improve the accuracy of object recognition by allowing mobile devices to contribute training samples in order to cater for different contexts during classification. More recently, authors in \cite{leidall2019edge} proposed an edge-based architecture where devices on the edge connected in P2P fashion run a semantic operating system to take the role of managing the sensors and actuators they own. IoT applications running on cloud or edge servers would then contact these devices to gain access to those sensors and actuators.

In this paper, in addition to the role of coordination, we utilize the edge server for execution of ambience intelligence tasks at the edge of network. A recent survey paper underscored the importance of this concept, named edge intelligence \cite{zhou2019edge}, and described challenges and future directions to observe it. In general, edge servers (a.k.a. cloudlets) provide high-bandwidth and low-latency access to resources needed to provide highly responsive services to mobile and IoT applications \cite{satyanarayanan2017emergence}. This is particularly important for the delay sensitive augmented and virtual reality applications. For example, an architecture involving edge servers was utilized to provide computation offloading for a cognitive assistance application for the elderly \cite{satyanarayanan2019augmenting}. This application performs object, face and text recognition on live images taken by a Goolge Glass device to offer guidance to users. Furthermore, edge servers were leveraged to run computer vision algorithms on live feeds of surveillance cameras \cite{yi2017lavea,zhang2015design}. Proposed use case includes automatically identifying people, objects or events of special interest to take necessary public safety measures.

%% file: design.tex
\section{Architecture}
\label{sec:design}
We design a context management framework that at its core benefits from the edge server as a central coordinator with sufficient computing capabilities to perform administrative work as well as accept offloaded tasks from mobile devices. An overview of the architecture of the framework is shown in Figure~\ref{fig:architecture}. Mainly, the framework design consists of software services running on edge server(s) and mobile devices. The software services perform the planning and coordination for the execution of ambience intelligence tasks across available assets. We use the term assets here to refer to both available IoT sensors in the environment as well as software components that can be used for performing intelligence tasks. For example, to perform object recognition, the needed assets are a camera facing the event to take an image and a software component such as a pre-trained Deep Neural Network (DNN) to perform inference on the image. In order to be able to devise efficient ambience intelligence execution plans, the framework gathers information about the available sensors in the environment. These sensors could be transiently available mobile sensors or tethered sensors that are installed as part of the IoT environment. We consider various ownership types and design the framework such that we benefit from any possible sensor that can be accessed to derive the context. The following sensor ownership schemes are supported by the framework:
\begin{figure}
\centering
\includegraphics[scale=0.6]{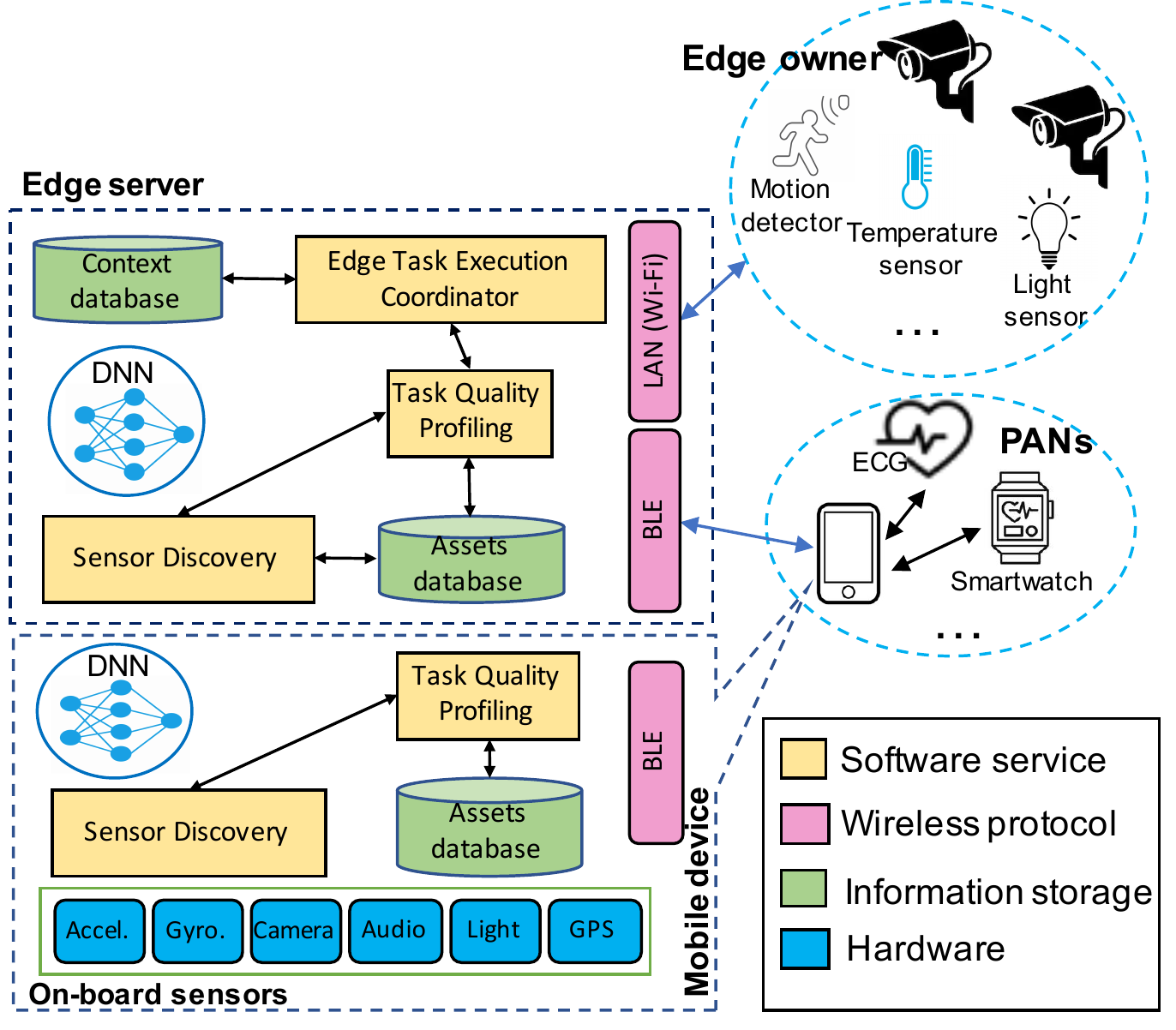}
\caption{Framework Architecture}
\label{fig:architecture}
\end{figure}
\begin{itemize}
\item \textbf{Mobile-connected sensors:} these sensors are either embedded inside the mobile device or connected to the mobile device via short range wireless protocols. Both types of sensors have the advantage of proximity to the sensor owner but are likely susceptible to inaccuracies of carrying positions. The framework considers both types to optimize for accuracy. As shown in the figure, our framework adapts the BLE protocol, which is the de facto communication protocol for the personal devices forming personal area network (PAN). As mentioned in the introduction, a key principle in our design is to allow the mobile device to negotiate and execute plans on behalf of the user. Hence, the mobile device will play the central role in the BLE communication by forming a star topology to form a bridge between the PAN devices on one side and the edge server on the other side. Consequently, the mobile device is capable of gathering information about all sensors on its side and on the edge side to select the sensor that is more suitable to perform the ambience intelligence task.
\item \textbf{Edge-connected sensors:} these sensors are installed in the environment and are likely owned by the edge deployment owner. Examples include a surveillance camera that also includes a microphone or a motion detector, temperature sensor or a light sensor. Those sensors are likely tethered, however, it is possible that they are not facing the event of interest directly or are far from the event. Hence, it is important to consider data from those sensors as well as from the mobile sensors during the planning phase to increase the chances of finding good quality sensor data to perform the ambience intelligence task.
\end{itemize}
The framework includes a {\it Sensor Discovery} module running on both the mobile device and the edge server. This module keeps a list of on-board sensors on the mobile device and adds to the list new sensors connected through BLE. On the edge server side, newly installed IoT sensors that are typically connected through Wi-Fi can be registered with the framework by the IoT owner. Sensors on both lists are polled periodically to check any disadvantageous factors and sensor information gets updated accordingly in the {\it Assets database} at both ends.
The polling process performs predefined simple checks that can mark a sensor as useless for a particular task. For example, a noisy environment for the audio sensor or a dark environment for the camera sensor. For edge servers the server can send heartbeat messages via {\it Wi-Fi interface} to check the status of the sensors.

In addition to checking sensor information, the framework includes a {\it Task Quality Profiling} service that is responsible for tracking performance metrics for the software components that perform intelligence tasks (i.e. {\it DNN Modules}). This service builds a model to predict the execution time for the software component given the input task size \cite{cuervo2010maui, alanezi2015panorama}, which is beneficial for time performance optimization decisions when deciding the distributed execution plan. In addition, any identifying information to execute the task along with performance metrics of execution accuracy are also stored in the {\it Assets database} to be used as part of the optimization process that is carried by the {\it Edge Task Execution Coordinator}. Note here that the mobile device, acting as a proxy for the user during the planning process exchanges information from its {\it Assets database} with the {\it Edge Task Execution Coordinator}. The latter uses the information gathered from the edge server and the mobile device to devise the best execution plan.
The framework also includes {\it Context database} where logs of discovered contexts that are calculated by the edge intelligence task are saved for retrieval by mobile devices.
Section \ref{sec:protocol} describes the protocol employed to collect assets information along with the type of the information collected. 

Finally, it is important to mention that we choose to implement the communication between the mobile device and the edge server to be carried using the Bluetooth Low Energy {\it BLE} protocol. This choice enables the mobile device to discover services on the edge server and exchange bursts of negotiation information with it retrieved from its own {\it Assets database} in an energy efficient manner as will be described in Section \ref{sec:evaluation}.

%% file: protocol.tex

\section{Protocol Design}
\label{sec:protocol}
\subsection{Functional Model}
\label{subsec:modell}
\begin{figure}
\centering
\includegraphics[scale=0.8]{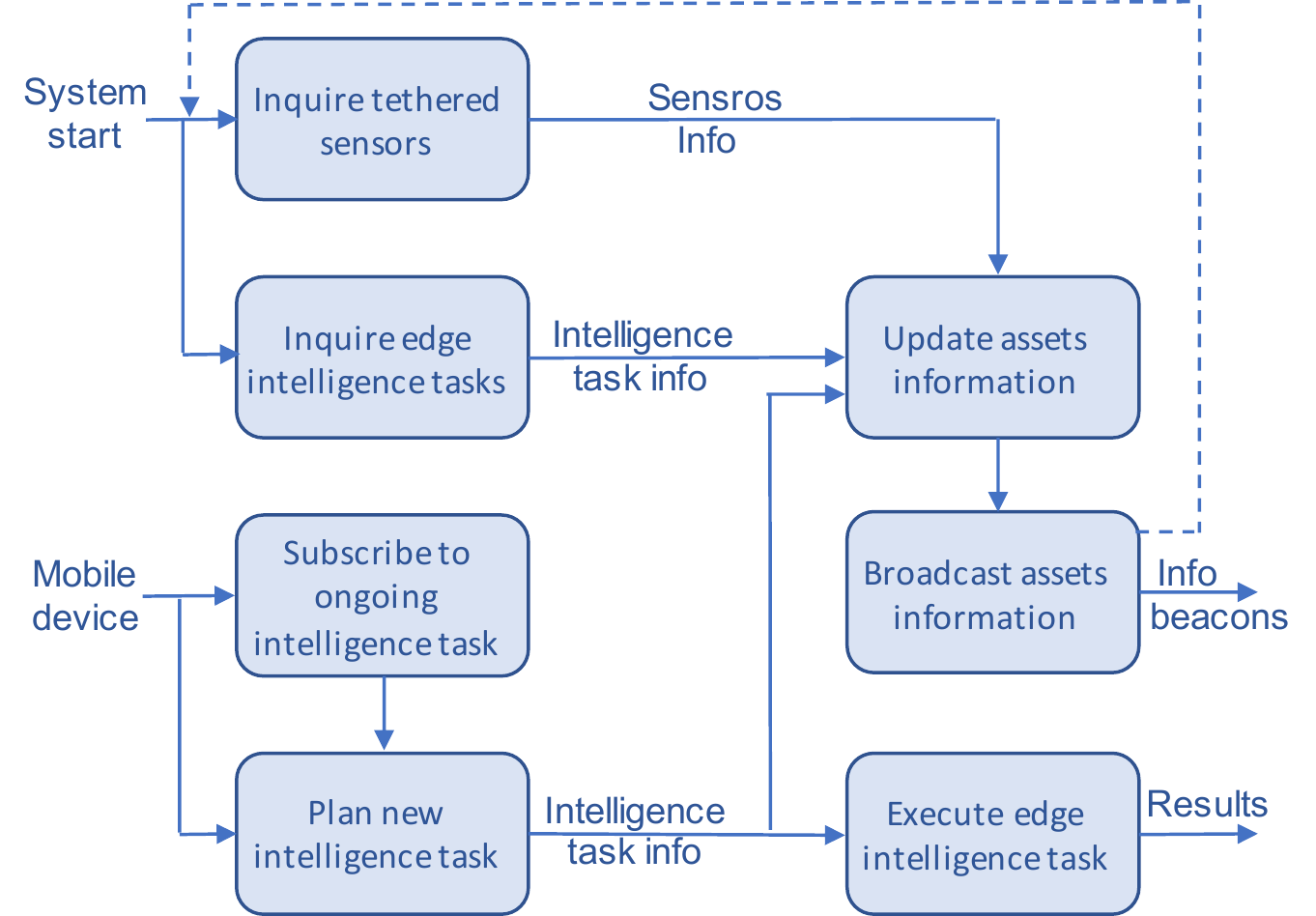}
\caption{Functional Model}
\label{fig:model}
\end{figure}
Performing the central role of coordination in an IoT environment requires the edge server to maintain updated information about all available assets. Consequently, this information could be used in conjunction with the information exchanged with user mobile devices to devise the most efficient plan according to the current situation. Figure~\ref{fig:model} depicts the functional model employed by the framework to be able to maintain environment information. The functional model describes the processes of the system and the flow of information between these processes. Upon starting the system, the framework running on the edge server sends inquiry messages to registered domain sensors ({\it Inquire tethered sensors}). Those sensors are sometimes tethered IoT sensors (i.e. connected to power supply) that are owned by the edge environment owner. In addition, they are likely connected with the edge server using WiFi. Examples include a surveillance camera or a temperature sensor. In addition, the framework performs an inquiry about available ambience intelligence software components that are installed on the edge server ({\it Inquire edge intelligence tasks}). This check helps the framework to track available services on the edge server in order to expose those services for discovery and reuse by mobile applications. Note here the wide range of ambience intelligence tasks such as computer vision, emotion analysis and sound analysis. This means that it is impractical to assume that they are all available on the edge server. Therefore, the framework employs a mechanism where it collects information about available services on the edge server to be used in the planning phase.

After collecting information for both sensors and edge intelligence tasks, the gathered information is passed for {\it updating assets information} in the assets database. Next, a summary of this information is encoded as service universally unique identifiers (UUIDs) in BLE advertisement packets that are {\it broadcast as assets information} in the BLE information beacons emitted by the edge server. This process of assets information discovery and broadcast is repeated periodically as indicated by the dashed arrow in order to help the edge server maintain current information about edge intelligence tasks and sensors. Information collected periodically about assets is described in details in Section \ref{subsec:assets}. We note that encoding assets as UUIDs in BLE beacons emitted by the edge server brings great energy savings to mobile devices when discovering edge environment services. BLE allows the mobile device to search for UUIDs of interest in the background. This means that the mobile device can search for a sensor (e.g. camera) or a service (e.g. face detection) while in sleep mode and only wake up when the desired service is discovered.   

Different components of the framework are also triggered based on mobile devices interaction with the system. When a mobile device discovers the edge server they can either {\it subscribe to an ongoing intelligence task} or invoke a new task. The availability of these tasks is learned from BLE beacons sent by the edge server. In case of a new task, the framework {\it plans the new intelligence task} using current assets information stored in the assets database in conjunction with assets information supplied by the mobile node to decide on the best execution plan. After that, the tasks information in the tasks database is updated to reflect that this task is currently running. Such updates are required in order to piggyback potential requests for the same intelligence task originating from other users. Subsequently, the execution time of an ongoing task requires merely reading the recent result of the task assuming that the result is not stale as per the  user application requirements. We reflect on time performance for executing ongoing and new tasks using two computer vision tasks, namely object and face recognition, as example ambience intelligence tasks in Section \ref{sec:evaluation}. Finally, the framework {\it executes the edge intelligence task} resulting from the plan and results are forwarded using BLE to requesting mobile device.

\subsection{Assets information}
\label{subsec:assets}
\begin{figure}
\centering
\includegraphics[scale=0.6]{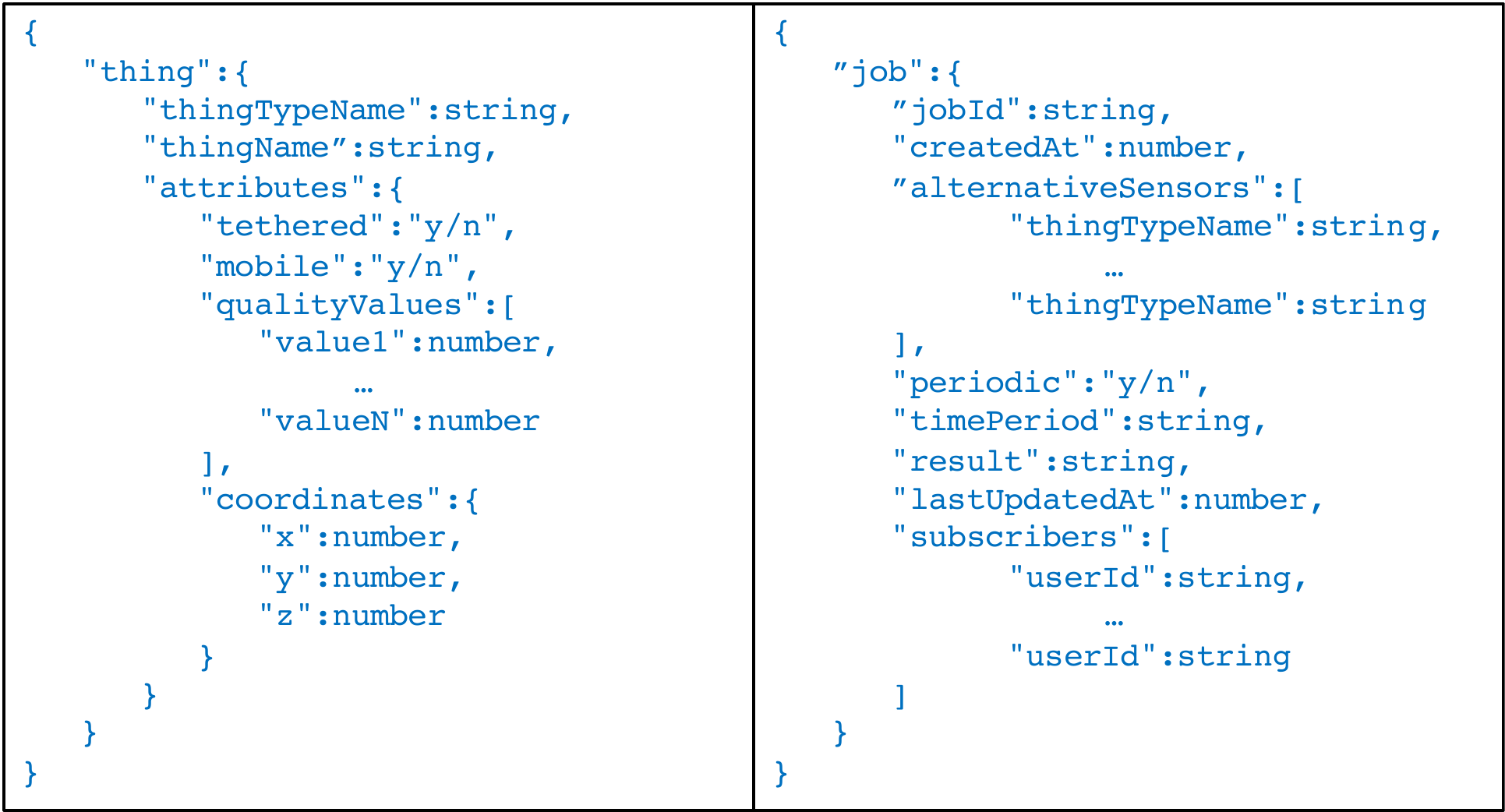}
\caption{Assets JSON Representation}
\label{fig:assets}
\end{figure}
The framework considers both the sensors and the ambience intelligence tasks' code as assets that are used when preparing the best execution plan. Hence, various features for these two types of assets must be gathered from the environment and exchanged during the planning phase. The JSON files containing the information of each asset type are shown in Figure~\ref{fig:assets}. Note here that choosing the right ontology to track and exchange assets information within an IoT network is related to the open research problem of interoperability in IoT. Various data formats are promoted in literature to tackle this problem \cite{noura2019interoperability} and we chose to adopt the Amazon AWS IoT standard \cite{aws-iot} with some additions to fit our design. Using this standard, devices that sense and act are called things while ambience intelligence tasks are represented by jobs.

The JSON file for things (i.e. sensor) is shown on the left. It contains identifying information about the sensor such as the type ({\it attribute: thingTypeName}) and the unique name chosen by the user ({\it attribute: thingName}). In addition, the framework tracks many features for the sensor that are grouped under {\it attributes}. First, it is important to know if the sensor is tethered, because energy saving for tethered sensors will not be of high priority. Conversely, mobile sensors are likely battery-powered and should be of lower priority for use so as to save energy. Furthermore, the JSON file contains multiple values related to the quality of the sensor that depend on the type of the sensor ({\it attribute: qualityValues}). For example, an image sensor can have a brightness value as a quality check for the ability of the sensor to be used at the moment for classification. On the other hand, volume or background noise can be the quality measures corresponding to audio sensors. The coordinates of the sensor can also be beneficial in case the coordinates of the measured event are known to assess the distance bewteen the sensor and the event, e.g. closer the sensor is to the event, better the quality it provides for the sensed data. 
Referring back to Figure~\ref{fig:assets}, the JSON file for the intelligence task (i.e. job in AWS IoT notation) is shown on the right. It lists multiple alternative sensors as possible sensors for a particular job ({\it attribute: alternativeSensors}). This helps the framework eliminate disadvantageous sensors based on quality checks while still being able to execute the task in hand using an alternative sensor with good quality values.
It is also important to track whether the job is periodic, the time period of the job and the subscribers. This information is used to repeat the job and send the results to subscribers. The timestamp of the last result from executing the job ({\it attribute: lastUpdatedAt}) is also tracked to allow subscribes to inspect the timeliness of the result.
\subsection{Coordination of Ambience Intelligence Task Execution}
\label{subsec:scenario}
This section provides an example of how the execution of an ambience intelligence is coordinated by the framework. Upon receiving a request for executing a task, the framework performs planning to decide the set of sensors and devices suitable for executing it. This decision is impacted by the current context. For example, let's consider a mobile application that requires discovering the identity of the people in a room. This task could possibly be executed in several different coordinated execution scenarios. One possible scenario could be to take a picture using the mobile device's camera and send it to the edge server for face recognition. Another scenario could be to access a nearby surveillance camera to get the needed picture. The first scenario could be problematic in case the smartphone is not in a good carry position (e.g. it is inside the user pocket). Hence, the framework would fall back to the surveillance camera access scenario. To select from these two scenarios the framework begins by inspecting the suitable sensors for the job listed in the job's JSON document described in Section \ref{subsec:assets}. Then, a quality check is performed for each sensor by comparing the current value (i.e. image brightness for the camera) with accepted quality value stated in the JSON document for the sensor (or thing). Accordingly, the camera sensor with the accepted quality is the one chosen for the job.

%% file: implementation.tex
\section{Experimental Evaluation}
\label{sec:evaluation}

\subsection{Implementation Prototype}

\begin{figure}
\centering
\includegraphics[scale=0.8]{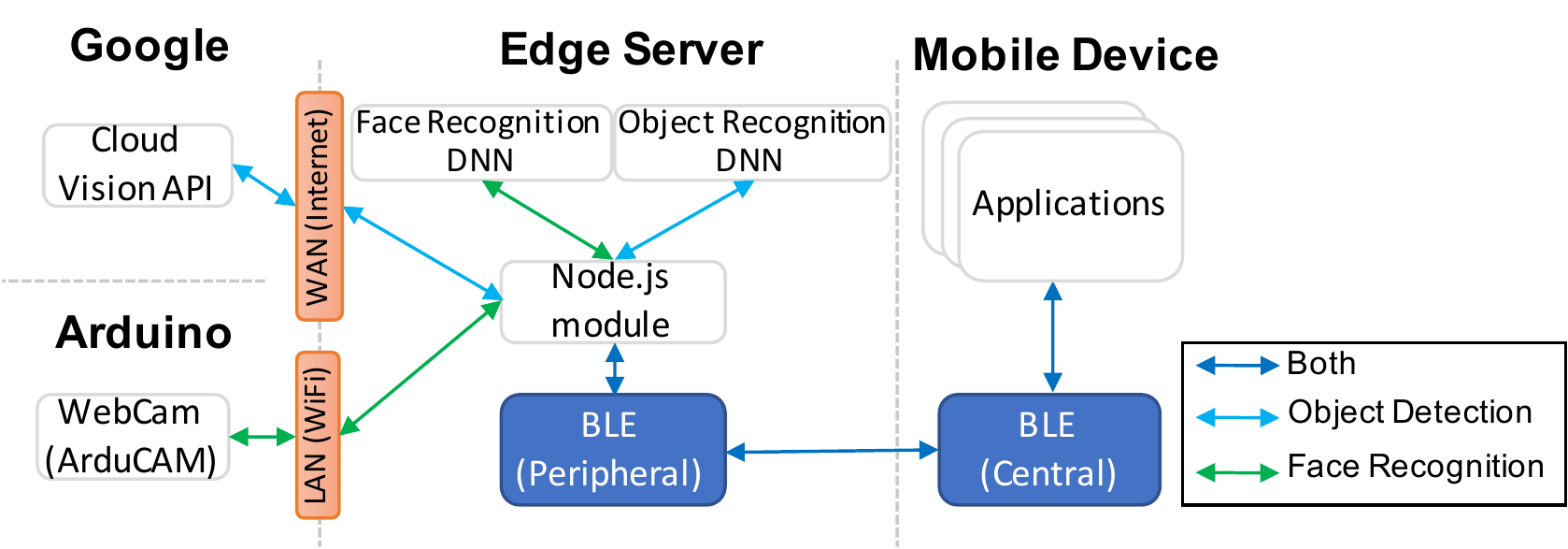}
\caption{Prototype Implementation}
\label{fig:prototype}
\end{figure}
Figure~\ref{fig:prototype} depicts a prototype that we implemented for the architecture described in Section \ref{sec:design}. In this prototype, we focus on the performance for the interaction between the edge server and mobile device, and between cloud and the camera, and use two computer vision tasks as example ambience intelligence tasks. Implementation of information gathering and planning is left out as a future work as we believe that it is worth its own full study.
A MacOS-based laptop is used to represent the edge server. The edge server is installed by the edge environment owner (e.g. house or business owner) to bring computation power and storage closer to the users with low latency when compared to accessing cloud resources.

The prototype includes an Android mobile device that acts as the proxy for the user by negotiating the execution of application intelligence task requests with the edge server without user intervention. Example of application intelligence tasks requests include emotion detection or speech recognition on sound clips, and face or object recognition on images or videos, which usually requires running compute intensive pre-trained models. We have implemented an Android client to discover and call the edge intelligence tasks on the server using BLE. For full implementation we run the framework code on the mobile device inside an Android service \cite{android-service}. This allows the framework to run in the background and accept delegation from mobile applications to execute ambience intelligence tasks. 
Finally, a 2MP Web Camera (ArduCAM ESP8266 UNO \cite{arducam}) is connected to the edge server via WiFi interface. An ESP2866 board is used to enable the microcontroller within the camera to communicate with the edge server over Wi-Fi. Table 1 describes the components of the prototype we have implemented.

\begin{table*}[htb]
\footnotesize
\label{Testbed}
\caption{Testbed}
\begin{center}
\begin{tabular}{ |c|c|c|c| } 
  \hline
 Title & Type & Function & Scenario Used \\
  \hline 
MacBook Air & Hardware & Edge computing server & Both \\
\hline
\thead{Motorola Moto \\E smartphone} & Hardware &  \thead{Mobile IoT device requesting \\services from the edge server} & Both \\
\hline
\thead{ArduCAM \\ESP8266 \cite{arducam}} & Hardware & Web camera connected to LAN &  \thead{Face \\Recognition} \\ 
\hline
Android Client & Software & \thead{Mobile user client requesting \\ambience intelligence tasks} & Both \\ 
\hline
Bleno \cite{bleno} & Software & \thead{Node.js library for implementing \\BLE peripheral on Mac OSX} & Both \\ 
\hline
\thead{Object Recognition \\DNN \cite{object-reco2}} & Software & \thead{DNN running on edge server to \\perform object recognition tasks} & \thead{Object \\Recognition} \\ 
\hline
\thead{Face Recognition \\DNN \cite{face-reco}} & Software & \thead{DNN running on edge server using \\Docker container to perform \\face recognition tasks} &  \thead{Face \\Recognition} \\ 
\hline
\thead{Google Cloud \\Vision API \cite{gc-vision}} & Software & \thead{Cloud service for performing object \\recognition tasks} & \thead{Object \\Recognition} \\ 
\hline

 \hline
\end{tabular}
\end{center}
\label{table:hardware}
\end{table*} 

As mentioned in Section \ref{sec:design}, the mobile device acts in BLE central role to discover and subscribe to the services of the edge server which acts in BLE peripheral role. We used bleno \cite{bleno}, which is written using Node.js to implement the BLE peripheral role on MacOS. Using this module, the edge server encodes the presence of its services in wireless broadcasts that can be heard by nearby mobile devices. It is worth noting that BLE allows devices in central mode (i.e. mobile devices) to scan for a particular service using its UUID while in sleep mode,
 thereby drastically minimizing the energy required for the detection of edge services.
The prototype utilized BLE 4 to enable this energy efficient discovery and messages exchange.  BLE 4 allows multiple central devices to connect with a peripheral device (i.e. the edge server), to coordinate and execute multiple services simultaneously. Our prototype only used BLE for messages exchange, hence the bandwidth limitations of BLE 4 were not applicable. The latest BLE 5 \cite{ble-5} improved on BLE 4 with double the bandwidth and 4-fold increase in communication range. This improvement can bring performance benefits for solutions that depend on BLE as in our framework.

The implementation of object recognition on the edge server uses Tensorflow \cite{abadi2016tensorflow} and is written using python. We used a pre-trained AlexNet model that has its weights stored in a file of size of 200MB. This file needs to be loaded in order to perform object recognition (i.e. inference) on images. As for face recognition, we ran a python-based face recognition docker image \cite{face-reco} on the edge server. This docker image provides API endpoints for adding faces to the database as well as to inquire about faces. We note here that there are many techniques in literature for optimizing performance (i.e. size, accuracy and execution time) of DNNs and these techniques are are orthogonal to our work.

\begin{figure*}[!t] 
\begin{minipage}[t]{0.45\textwidth}
\includegraphics[width=\textwidth]{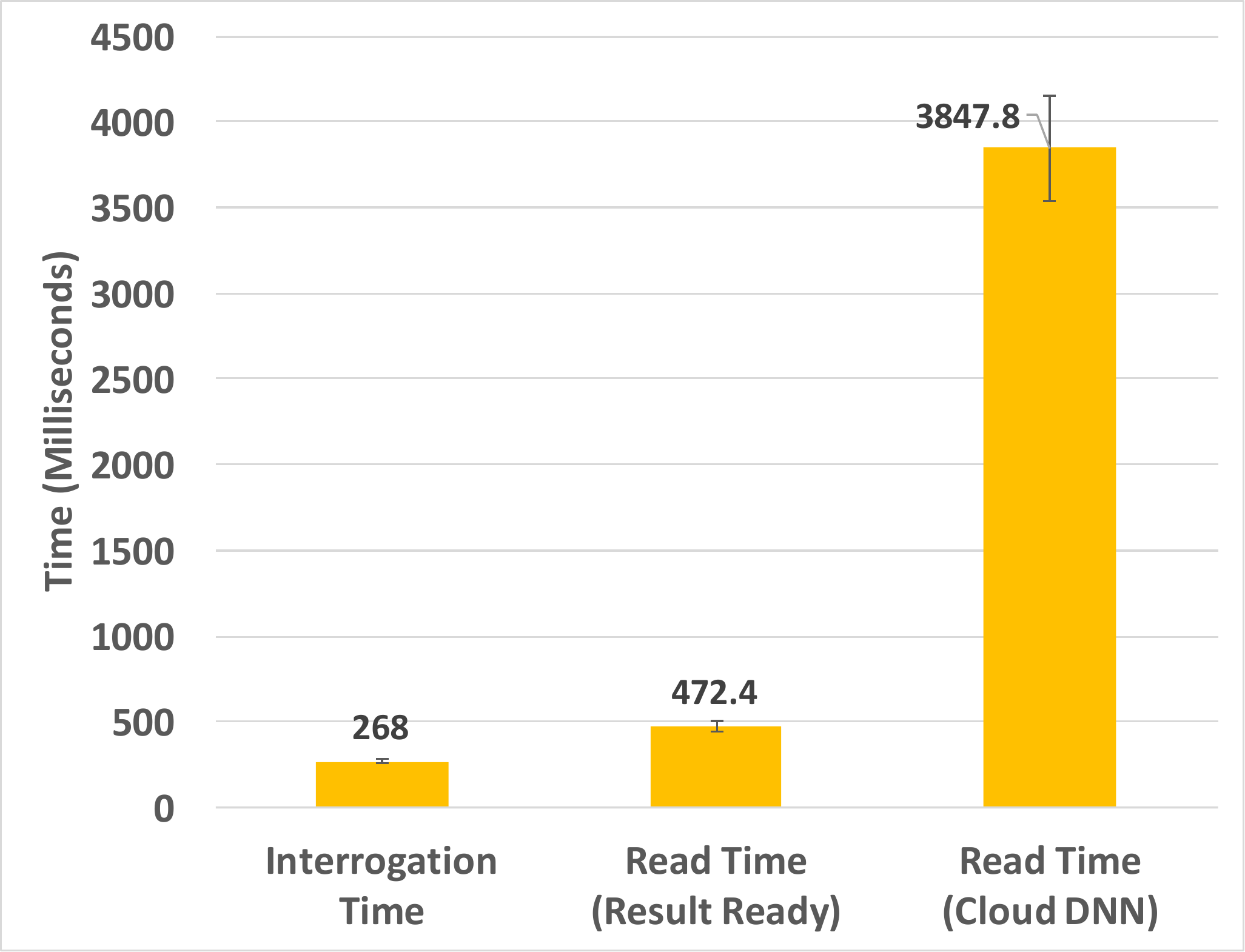} 
\caption{Time measurements for edge server interrogation and result reading time for the situations when the object recognition result is available or not available on the edge.}
\label{fig:results} 
\end{minipage} \hfill
\begin{minipage}[t]{0.45\textwidth}
\includegraphics[width=\textwidth]{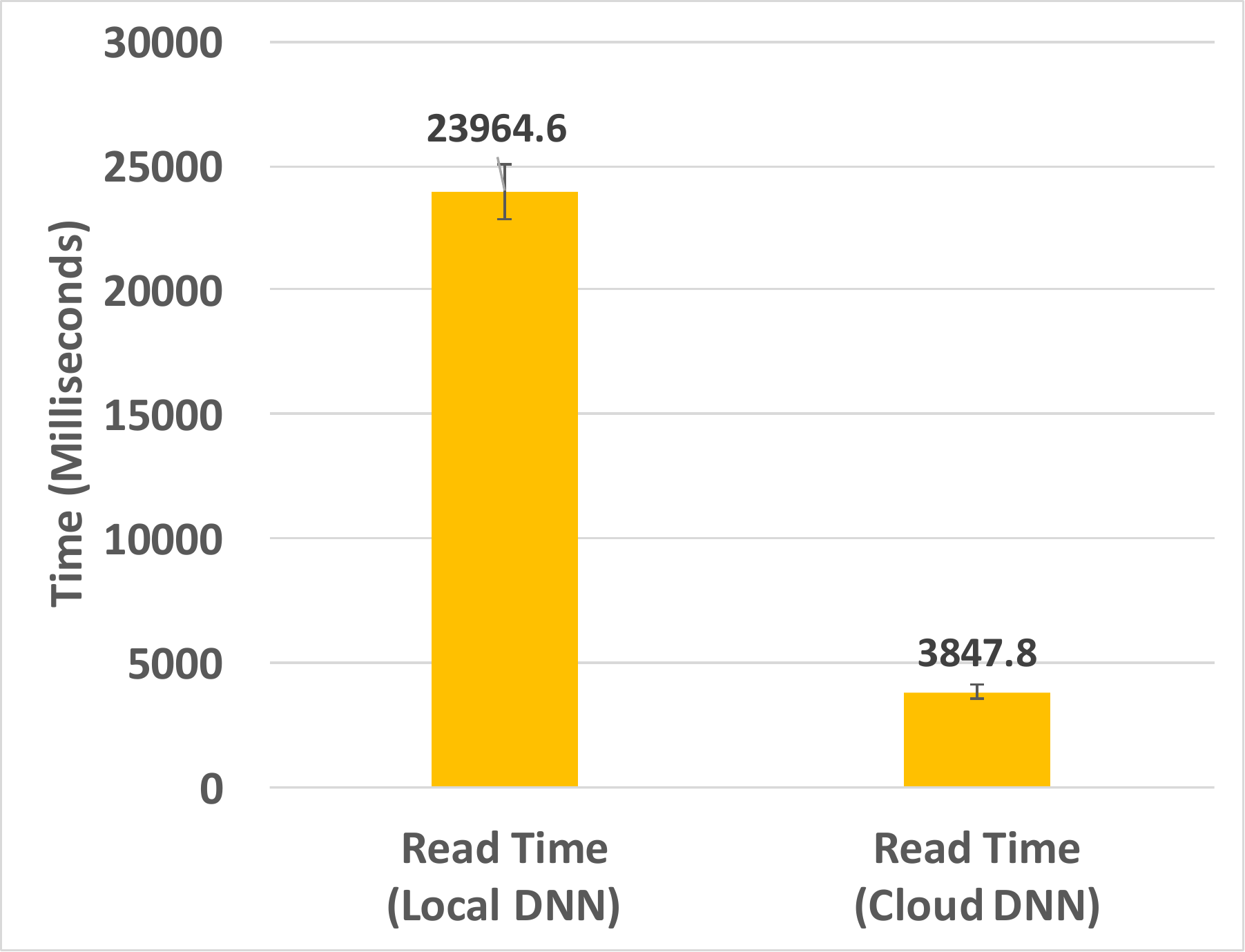} 
\caption{Time measurements for reading object recognition results from the edge server when the server has access to GC vision API vs loading and executing a locally stored model.}
\label{fig:not-ready} 
\end{minipage} \hfill
\end{figure*}

As we can see, in the prototype we implemented, the edge server has three interfaces. The first interface is with the client over BLE with which a client discovers and subscribes to edge services, the second interface is with the sensors in the environment over WiFi, and the third interface is with cloud. We have conducted several experiments to evaluate the performance of each of these interfaces as discussed below.

\subsection{Edge Server - Client Interaction}
In our first experiment, we used object recognition on images stored on the edge server as the ambience intelligence task. This generic scenario resembles a mobile application interested in receiving tags naming individual objects identified from an image. Similar to object recognition, we envision wide range of ambience intelligence tasks to be deployed on edge servers to support the edge intelligence vision \cite{zhou2019edge}. In this scenario, images are assumed to be available on the edge server, which is quite reasonable assuming that a surveillance camera is periodically uploading images to the edge server for analytics including object recognition. There are multiple plausible scenarios covered in our measurements in Figure~\ref{fig:results} and Figure~\ref{fig:not-ready} in regards to the status of the object recognition result on the edge server. The first scenario occurs when the timestamp of the result of the latest object recognition is acceptable for the requesting mobile application (result ready). In case this is not true, the framework can either request to perform object recognition locally (local DNN) or upload the image to Google cloud vision API \cite{gc-vision} (cloud DNN) depending on the availability of Internet access to reach the APIs.

Since the prototype we implemented is distributed in nature, we focus in our evaluation on the time performance for executing these different scenarios. First, we report the time delay in executing two milestones for interactions between the edge server and the mobile device. The two milestones are the interrogation time and the result reading time. The interrogation time is the time elapsed between discovering the presence of the edge server from the broadcasts to the time the mobile device is ready to invoke the services offered by the edge server. This time is related to the mechanics of how BLE works in which services and their characteristics must be learned before communicating through them.
On the other hand, the reading time includes the interrogation time plus the time needed to finally receive the object recognition result. Both of these timings are measured from the mobile device (i.e. user) side. We report in each experiment the average results from running the same experiment five times along with the standard error shown on the bars.

We see in Figure~\ref{fig:results} that the interrogation time takes an average of 260 milliseconds. Whereas, the read time when the result is ready on the edge is 472 milliseconds. Since we report aggregate times, this result means that the read time also includes the interrogation time. The object recognition code stores the fresh result obtained periodically in a text file and the read time is the time required to open the file and encode the result in the response for the read request issued through BLE by the mobile device.

\subsection{Edge Server - Cloud Interaction}
If the mobile client doesn't accept the object recognition result after checking the timestamp, an image recognition request is sent by the edge server to Google cloud vision API to obtain the object recognition result on a latest image available on the edge server. The average time required to obtain object recognition results from the API and report it back to the mobile device is 3847 milliseconds (See Figure~\ref{fig:results}). 

\begin{figure*}[!t] 
\begin{minipage}[t]{0.45\textwidth}
\includegraphics[width=\textwidth]{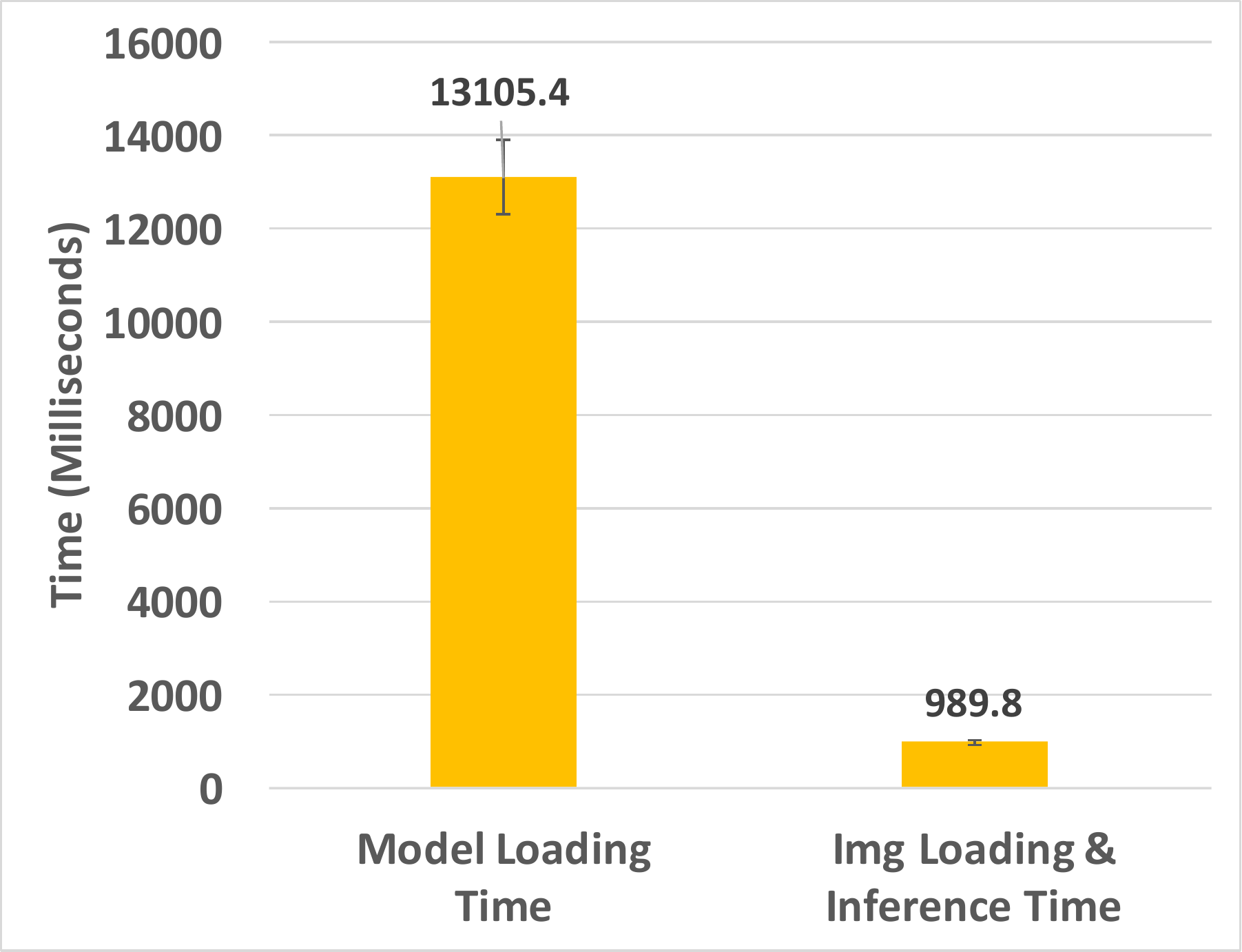} 
\caption{Time measurements for the stages of loading and executing a pre-trained object recognition model (Local DNN).}
\label{fig:pretrained} 
\end{minipage} \hfill
\begin{minipage}[t]{0.45\textwidth}
\includegraphics[width=\textwidth]{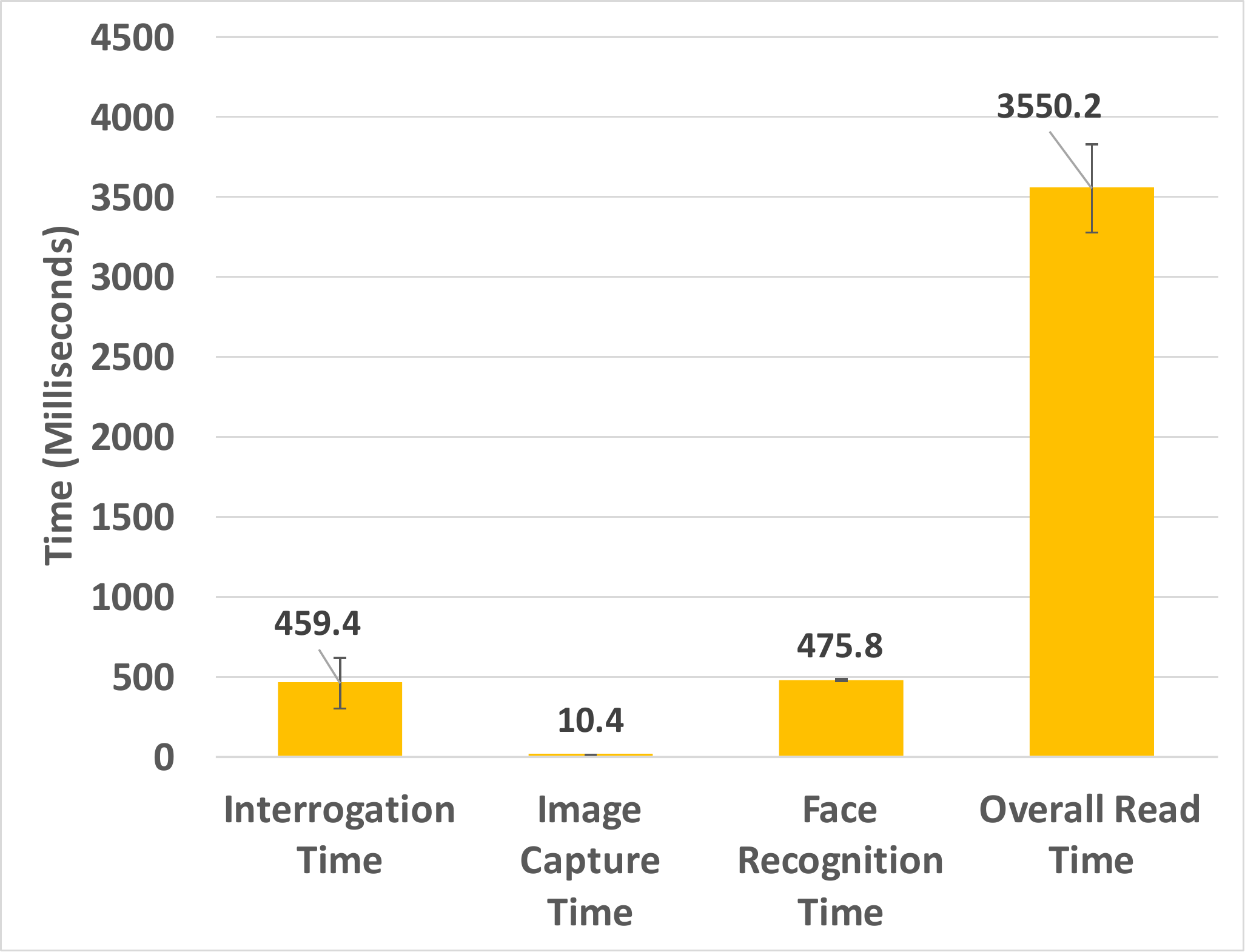} 
\caption{Time measurements for the various stages involved in the face recognition scenario.}
\label{fig:face-time}  
\end{minipage} \hfill
\end{figure*}

Another option other than calling the Google cloud vision API is to use a locally pre-trained model that is stored on the edge server. We report in Figure~\ref{fig:not-ready} a comparison between calling the Google cloud vision API against loading and executing the local model. The average time for loading and executing the pre-trained model and returning the result to the mobile client is 23964 milliseconds. Needless to say, the delays involved in the process of loading and executing the pre-trained model will not be tolerated by clients and the solution to this problem is to preload the pre-trained model and use it for inference whenever required. To reflect to the reader the time penalty when only inference is required, we report the major milestones for loading and executing the pre-trained model in Figure~\ref{fig:pretrained}. We notice from the figure that this process is dominated by the model loading time which is expected due to the huge size of the file storing the model weights at around 200MB. However, the image loading time plus inference time is only 989 milliseconds. Hence, when we combine the results from Figure~\ref{fig:not-ready} and Figure~\ref{fig:pretrained}, we see that the best option in case the latest object recognition result available on the server is expired is to call a preloaded pre-trained model, which will add a performance penalty of less than a second (i.e. inference only). However, this approach requires preloading various pre-trained DNN models pertaining to different ambience intelligence tasks to the edge server memory to be ready to serve client requests for ambience intelligence tasks.

\begin{figure*}[!t] 
\begin{minipage}[t]{0.45\textwidth}
\includegraphics[width=\textwidth]{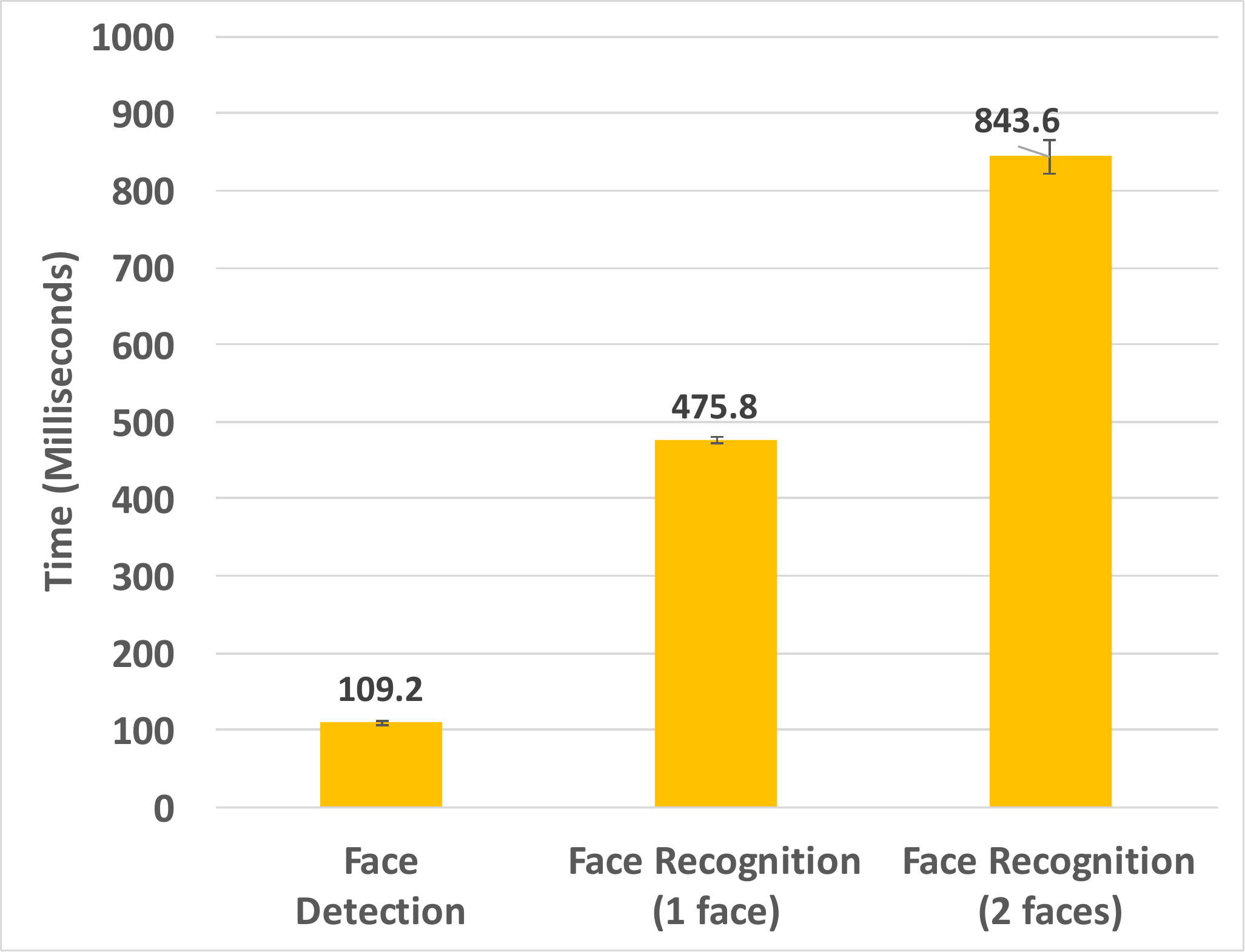} 
\caption{Time performance for face recognition when performing face detection, face recognition for a single face, and face recognition for two faces using a pre-loaded face recognition service on the edge.}
\label{fig:face-no} 
\end{minipage} \hfill
\begin{minipage}[t]{0.45\textwidth}
\includegraphics[width=\textwidth]{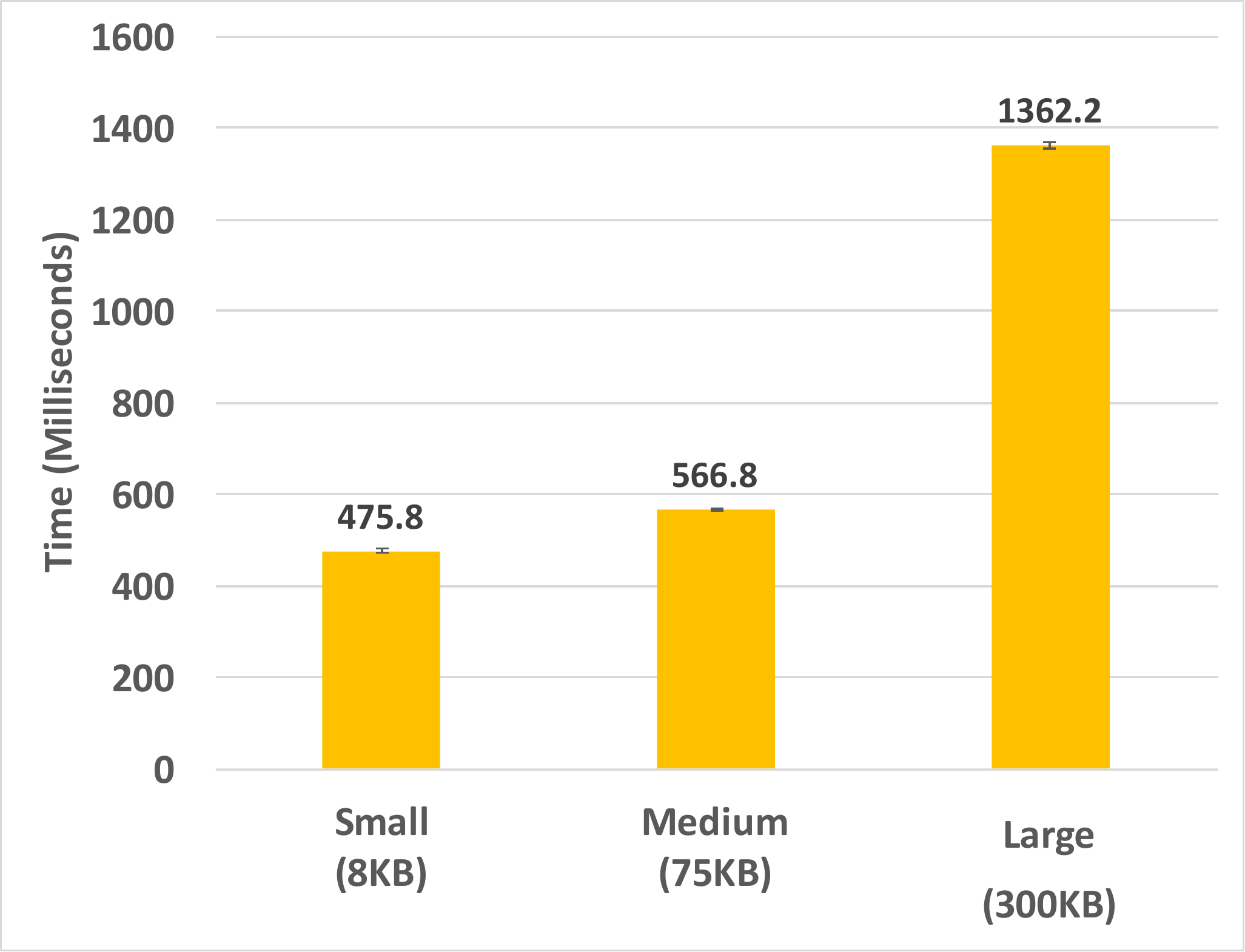} 
\caption{Time performance for face recognition when processing a small, medium, and large size files using a pre-loaded face recognition service on the edge.}
\label{fig:face-size} 
\end{minipage} \hfill
\end{figure*}

\subsection{Edge Server - Camera Interaction}

We now look at the performance when the edge server interacts with the camera. Unlike the previous experiment, we do not assume that images are already available on the edge server. Instead, on receiving a service request, the edge server interacts with the camera and retrieves an image on which it performs face recognition. This experiment consists of the client first discovering and subscribing to the services of the edge server via BLE, the edge server capturing an image from the camera over Wi-Fi, and then performing face recognition locally and returning a result to the client. Figure~\ref{fig:face-time} shows the performance for various stages involved in this experiment. First, the interrogation time to discover and subscribe to the edge service took 459.4 milliseconds. Image capture time from the camera over WiFi was very fast. It took only 10.4 milliseconds to report the image to the edge server from the camera. In this experiment, the captured image had only one face, and it took an average of 475.8 milliseconds for the face recognition to recognize this face and report the identity of the person in the image. Finally, the overall read time for receiving the face recognition result back at the requesting mobile client over BLE is 3550.2 milliseconds.

In order to gain insight into how the number of faces in the captured image and image sizes impact the performance, we repeated this experiment for two different scenarios. First, we increased the number of faces in the captured image. Figure~\ref{fig:face-no} shows the performance when there are one face and two faces in the image respectively reported alongside the situation when there are no faces in the image (i.e. only face detection algorithm is executed). As we can see, the number of faces in the image increases the recognition time in which each new face adds about 360 milliseconds after face detection. Figure~\ref{fig:face-size} shows the impact of image size. As we can see, as the image resolution and thereby the image size increases, the face recognition time increases as well. We utilized the default image resolution of the camera, which reported a small image of an average size of 8KB only. As can be seen in the figure, the time it takes to perform face recognition on this image when there is one face is 475.8 milliseconds. However, assuming that the area to be monitored for face recognition is wider, higher image resolutions are required. From the same figure we can see that when the image resolution is increased to 75KB (10x), the time it takes to recognize the face increases by around 90 milliseconds. Increasing the image size (i.e. resolution) furthermore to 300KB leads to significant increase in the face recognition time to 1.3 seconds. We learn from these results that it is vital that the distributed edge services must be designed keeping time efficiency in mind in order to preevnt any negative impact on the responsiveness of requesting applications.

%% file: conclusion.tex
%
%
%
%
%
%
\section{Conclusion and Future Work}
This paper presents an architecture that is based on edge computing and the IoP paradigm to devise efficient collaboration plans to execute ambience intelligence tasks. Using BLE, mobile devices discover services on the edge server and exchange necessary information about their capability. The edge server utilizes this information along with other information about its capability and available sensors in the environment in the planning process. We implemented a prototype of the architecture using object recognition and face recognition as two examples of intelligence tasks. Time measurements demonstrate that the cost of inference for these tasks is quite good. We plan as a future work to perform full implementation of the architecture and involve PAN devices in the picture to capture performance metrics related to periodic information gathering and the planning phases.